\documentclass{ws-p8-50x6-00}

\newcommand{\AmS}{{\protect\the\textfont2
  A\kern-.1667em\lower.5ex\hbox{M}\kern-.125emS}}

\def \gsim {~\mbox{${}^> \hspace*{-9pt} _\sim$}~}

\def \znbb {$0\nu\beta\beta$}

\newcommand{\ba}[1]{\begin{eqnarray} \label{(#1)}}
\newcommand{\ea}{\end{eqnarray}}

\begin{document} 

\title{Neutrinoless Double Beta Decay - \hspace{15.cm} 
	And The Neutrino Mass Matrix}

\author{H.V. Klapdor-Kleingrothaus}

\address{Max-Planck-Institut f\"ur Kernphysik,\\ 
P.O. Box 10 39 80, D-69029 Heidelberg, Germany\\ 
Spokesman of HEIDELBERG-MOSCOW and GENIUS Collaborations\\
E-mail: klapdor@gustav.mpi-hd.mpg,\\
 Home-page: http://www.mpi-hd.mpg.de.non\_acc/}

\maketitle


\abstracts{
	Double beta decay is indispensable to solve the question of 
	the neutrino mass matrix together with $\nu$ oscillation experiments. 
	Recent analysis of the 
	most sensitive experiment since eight years - 
	the HEIDELBERG-MOSCOW experiment in Gran-Sasso - 
	yields evidence for the neutrinoless decay mode. 
	This result is 
	the first evidence for lepton number violation and proves 
	the neutrino to be a Majorana particle. 
	We give the present status of the analysis 
	in these Proceedings.
	It excludes several of the neutrino mass scenarios 
	allowed from present neutrino oscillation experiments - 
	essentially only degenerate and partially degenerate mass 
	scenarios survive. This result allows neutrinos to still play 
	an important role as dark matter in the Universe.
	To improve the present result, considerably enlarged 
	experiments are required, such as GENIUS. 
	A GENIUS Test Facility has just been funded and will 
	 come into operation by end of 2002.}


\section{Introduction}
	Recently atmospheric and solar neutrino oscillation experiments 
	have shown that neutrinos are massive. This was the first 
	indication of beyond standard model physics. The absolute 
	neutrino mass scale was, however, still unknown, 
	and only neutrino oscillations and neutrinoless double beta decay 
	{\it together} can solve this problem (see, e.g. 
\cite{KKPS,KKPS-01,KK60Y}). 
	Another question of fundamental importance is, 
	that whether the neutrino is a Dirac or Majorana particle. 
	Only double beta decay can solve this problem. 
	Finally double beta decay has a unique sensitivity 
	to probe lepton number conservation.

	In this paper we will discuss the status of double 
	beta decay search. 
	We shall, in section 2, 
	discuss the expectations for the observable of neutrinoless double 
	beta decay, the effective neutrino mass $\langle m_\nu \rangle$, 
	from the most recent $\nu$ oscillation experiments.  
	In section 3 we shall discuss the 
	recent evidence for the neutrinoless decay mode,  
	and discuss the consequences for the neutrino mass scenarios 
	which could be realized in nature.
	In section 4 we discuss the 
	possible future potential of $0\nu\beta\beta$ experiments,
	which could improve the present accuracy. 
	


\section{Allowed Ranges of 
$\langle m \rangle$ by $\nu$ Oscillation Experiments}

	  The obser\-vable of double beta decay 

\centerline{$\langle m \rangle =
|\sum U^2_{ei}m^{}_i| = |m^{(1)}_{ee}| 
		      + e^{i\phi_2} |m^{(2)}_{ee}| 
		      + e^{i\phi_3} |m^{(3)}_{ee}|,
$}

\noindent
	  with $U^{}_{ei}$ 
	  denoting elements of the neutrino mixing matrix, 
	  $m_i$ neutrino mass eigenstates, and $\phi_i$  relative Majorana 
	  CP phases, can be written in terms of oscillation parameters 
\cite{KKPS,KKPS-01} 
\begin{eqnarray}
\label{1}
|m^{(1)}_{ee}| &=& |U^{}_{e1}|^2 m^{}_1,\\
\label{2}
|m^{(2)}_{ee}| &=& |U^{}_{e2}|^2 \sqrt{\Delta m^2_{21} + m^{2}_1},\\
\label{3}
|m^{(3)}_{ee}| &=& |U^{}_{e3}|^2 \sqrt{\Delta m^2_{32} 
				 + \Delta m^2_{21} + m^{2}_1}.
\end{eqnarray}

	The effective mass $\langle m \rangle$ is related with the 
	half-life for $0\nu\beta\beta$ decay via 
$\left(T^{0\nu}_{1/2}\right)^{-1}\sim \langle m_\nu \rangle^2$, 
        and for the limit on  $T^{0\nu}_{1/2}$
	deducible in an experiment we have 
	$T^{0\nu}_{1/2} \sim a \sqrt{\frac{Mt}{\Delta E B}}$.
	Here $a$ is the isotopical abundance of the $\beta\beta$ emitter;
	$M$ is the active detector mass; 
	$t$ is the measuring time; 
	$\Delta E$ is the energy resolution; 
	$B$ is the background count rate. 

%
 \vspace{-0.5cm}
\begin{figure}[ht]
\vspace{9pt}
\centering{
\includegraphics*[scale=0.45]
{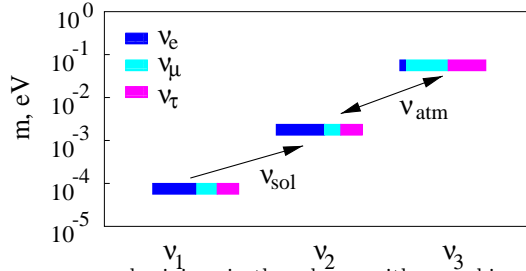}
\vspace{-0.5cm}
\caption[]{
       Neutrino masses and mixings in the scheme with mass hierarchy. 
       Coloured bars correspond to flavor admixtures in the mass 
       eigenstates $\nu_1, \nu_2, \nu_3$. 
       The quantity $\langle m \rangle$
	 is determined by the dark blue bars denoting 
	 the admixture of the electron neutrino $U_{ei}$.
\label{fig:Hierarchi-NuMass}}}
\end{figure}
%

	Neutrino oscillation experiments fix or restrict some of the 
	parameters in 
(1)--(3), e.g. in the case of normal hierarchy solar neutrino 
	  experiments yield 
	  $\Delta m^2_{21}$, 
	  $|U_{e1}|^2 = \cos^2\theta_{\odot}$ 
	  and
	  $|U_{e2}|^2 = \sin^2\theta_{\odot}$. 
	  Atmospheric neutrinos fix  
	  $\Delta m^2_{32}$, 
	  and experiments like CHOOZ, looking for $\nu_e$ 
	  disappearance restrict $|U_{e3}|^2$. 
	  The phases $\phi_i$  and the mass of the lightest neutrino, 
	  $m_1$ are free parameters. 
	Double beta decay can fix the parameter m$_1$ and thus 
	the absolute mass scale.
	  The expectations for 
$\langle m \rangle$ 
	  from oscillation experiments in different neutrino mass scenarios 
	  have been carefully analyzed in  
\cite{KKPS,KKPS-01}. In sections 2.1 to 2.3 we give some examples.



\begin{figure}[ht]
\centering{\includegraphics*[scale=0.40]{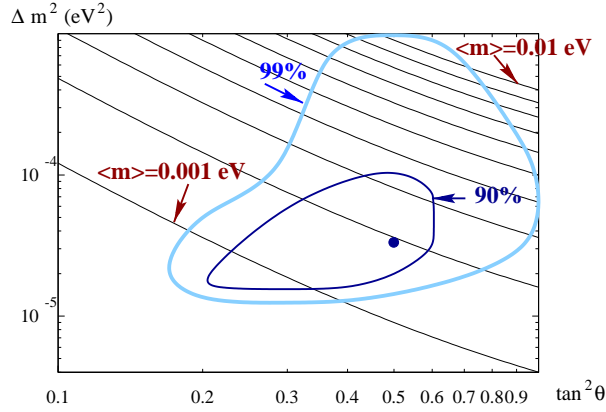}}
\caption[]{
       Double beta decay observable
$\langle m \rangle$
       and oscillation parameters in the case of the MSW 
       large mixing angle solution of the solar neutrino deficit, 
       where the dominant contribution to $\langle m \rangle$ comes 
       from  the  second  state. 
       Shown are  lines of constant $\langle m \rangle$,  
       the lowest  line corresponding to 
       $\langle m_\nu \rangle = 0.001$~eV, 
       the upper line to 0.01~eV. 
       The inner and outer closed line show the regions allowed 
       by present solar neutrino experiments with 
	90\% C.L. and 99\% C.L., 
       respectively. 
       Double beta decay with sufficient sensitivity could check the 
       LMA MSW solution. 
       Complementary information could be obtained from the search for a 
       day-night effect and spectral distortions in future solar 
       neutrino experiments as well as a disappearance signal in KAMLAND 
	[from \cite{KKPS-01}]. 
\label{fig:Dark2}
}
\vspace{.5cm}
\end{figure}



\subsection{\it Hierarchical Spectrum $(m_1 \ll m_2 \ll  m_3)$}
         In hierarchical spectra
(Fig.~\ref{fig:Hierarchi-NuMass}), 
	motivated by analogies with the quark sector and the simplest 
	  see-saw models, the main contribution comes from 
	  $m_2$ or $m_3$. 
	  For the large mixing angle (LMA) MSW solution which is favored 
	  at present for the solar neutrino problem (see 
\cite{Suz00}), the contribution of $m_2$ becomes dominant in the expression 
       for $\langle m \rangle$, and  
\begin{equation}
\langle m \rangle \simeq m^{(2)}_{ee} 
	= \frac{\tan^2\theta}{1+\tan^2 \theta}\sqrt{\Delta m^2_{\odot}}.
\end{equation}
	In the region allowed at 90\% C.L. by Superkamiokande according to 
\cite{Val01}, 
	the prediction for $\langle m \rangle$ becomes        
\begin{equation}
\langle m \rangle =(1\div 3) \cdot 10^{-3} {\rm eV}.
\end{equation}
	The prediction extends to 
	$\langle m \rangle = 10^{-2}$ eV in the 99\% C.L. range 
(Fig. ~\ref{fig:Dark2}).


\begin{figure}[h]
\centering{
\includegraphics[scale=0.45]
{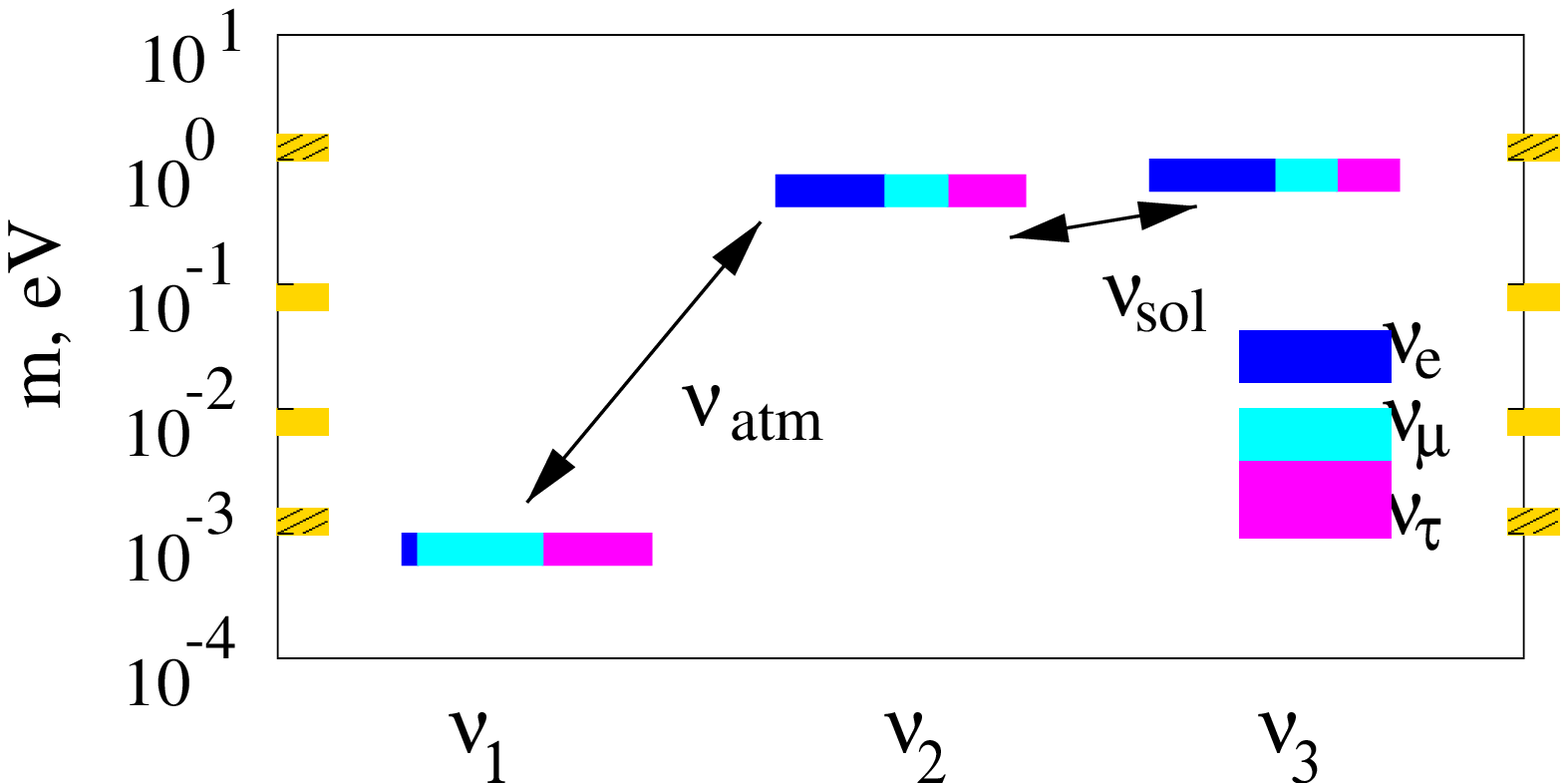}
\hspace{2.cm}
\includegraphics[scale=0.45]
{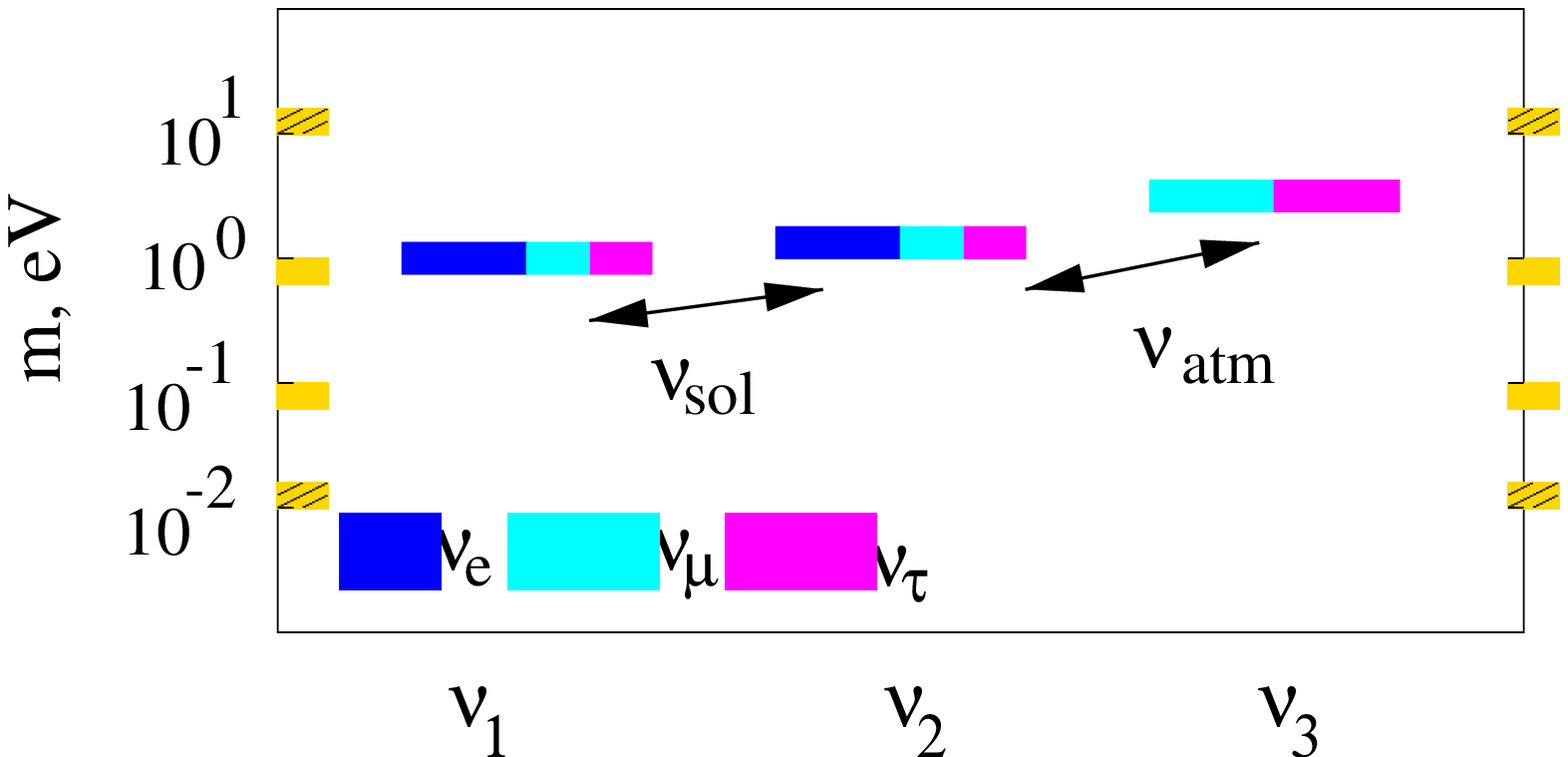}
}
\caption[]{
       \underline{Left:} Neutrino masses and mixing in the inverse 
	hierarchy scenario. \underline{Right:} Neutrino masses and mixings 
	in the degenerate scheme.
\label{fig:INVERSE-NuMass}
\label{fig:Degener-NuMass}}
\end{figure}



\subsection{\it Inverse Hierarchy $(m_3 \approx m_2 \gg  m_1)$}
           In inverse hierarchy scenarios 
(Fig.~\ref{fig:INVERSE-NuMass})   
	the heaviest state with mass $m_3$ is mainly the electron 
	   neutrino, its mass being determined by atmospheric neutrinos, 
$m_3 \simeq \sqrt{\Delta m^2_{\rm atm}}$.
	   For the LMA MSW solution one finds 
\cite{KKPS-01}
\begin{equation}
\langle m \rangle 
= (1\div 7) \cdot 10^{-2} {\rm eV}.
\end{equation}



\subsection{\it Degenerate Spectrum $(m_1 \simeq m_2 \simeq m_3 
	\gsim	 0.1~$eV)}
        In degenerate scenarios (fig. \ref{fig:Degener-NuMass}) the 
	contribution of $m_3$ is strongly restricted by CHOOZ.  
	The main contributions come from $m_1$ and $m_2$, depending on 
	their admixture to the electron flavors, which is determined 
	by the solar neutrino solution. We find 
\cite{KKPS-01}
\begin{equation}
m_{\min} < \langle m \rangle < m_1 \qquad 
\mbox{with} \qquad 
\langle m_{\min}\rangle = 
	(\cos^2\theta_{\odot} -\sin^2\theta_{\odot})\, m^{}_1.
\end{equation}

       	This leads for the LMA solution to 
$\langle m \rangle = (0.25\div 1)\cdot m_1$, 
	 the allowed range corresponding to possible values of 
	 the unknown Majorana CP-phases.

	 After these examples we give a summary of our analysis 
\cite{KKPS,KKPS-01} 
	of the $\langle m \rangle $ allowed by $\nu$ oscillation 
      experiments for neutrino mass models in the presently 
      favored scenarios, 
in Fig.~\ref{fig:Jahr00-Sum-difSchemNeutr}. 
	  The size of the bars corresponds to the uncertainty in 
	  mixing angles and the unknown Majorana CP-phases.



\begin{figure}[h]
\centering{
\includegraphics*[scale=0.45]
{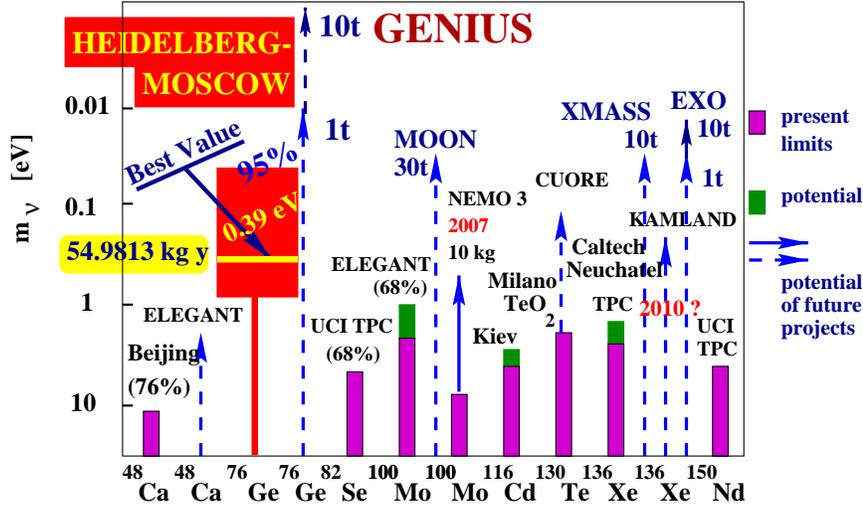}}
\caption[]{
       Present sensitivity, and expectation for the future, 
       of the most promising $\beta\beta$ experiments. 
       Given are limits for $\langle m \rangle $, except 
	for the HEIDELBERG-MOSCOW experiment where the recently 
	observed {\it value} is given (95$\%$ c.l. range), and the best value.
	Framed parts of the bars: present status; not framed parts: 
       expectation for running experiments; solid and dashed lines: 
       experiments under construction or proposed, respectively. 
       For references see 
\protect\cite{KK60Y,KK02,KK-LowNu2,KK-NANPino00}.
\label{fig:Now4-gist-mass}}
\end{figure}



\section{Evidence for the Neutrinoless Decay Mode}

	The status of present double beta experiments is shown in 
Fig.~\ref{fig:Now4-gist-mass}	
	and is extensively discussed in 
\cite{KK60Y}.	
	The HEIDELBERG-MOSCOW experiment using the largest source strength 
	of 11 kg of enriched $^{76}$Ge in form of five HP Ge-detectors 
	is running since August 1990 
	in the Gran-Sasso underground laboratory 
\cite{KK60Y,HDM01,KK02,KK-StProc00,HDM97}, 
	and is since long time the most sensitive one. 
	We communicate here the status of the analysis 
	of November 2001.	

\subsection{\it Data from the HEIDELBERG-MOSCOW Experiment}

\begin{figure}[h]
\centering{
\includegraphics*[scale=0.30]
{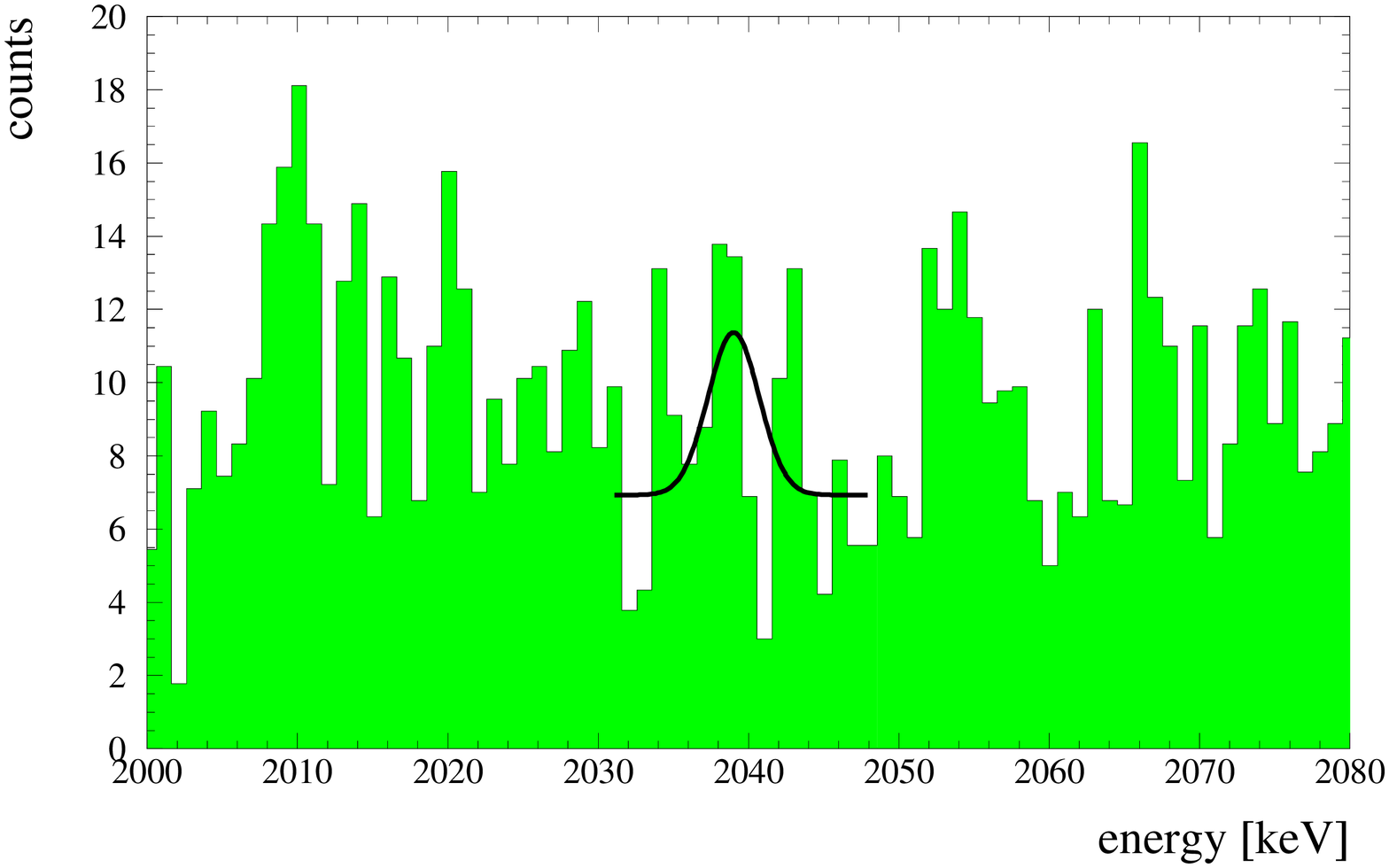} 
\includegraphics*[scale=0.30]
{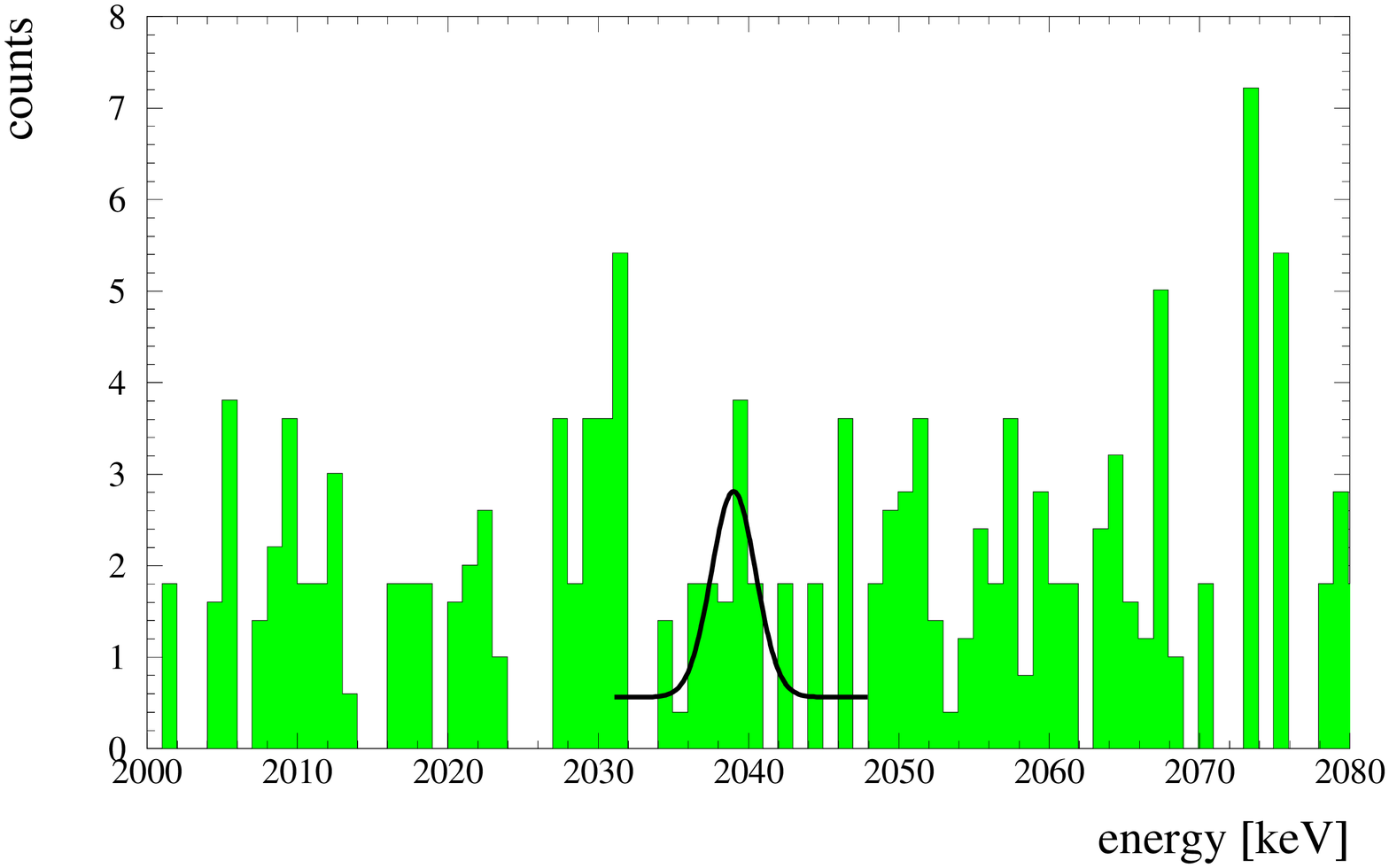} 
}

\vspace{-.5cm}
\caption[]{
       HEIDELBERG-MOSCOW experiment. Left: energy spectrum in the range 
	between 2000 keV and 2080 keV, 
	after 54.9813\,kg\,y.   
	Q$_{\beta\beta}$ = 2039.006(50) keV according to 
\cite{New-Q-2001}.
	Right: Spectrum of single site events 
	(SSE) after 28.053\,kg\,y, corrected 
	for the efficiency of SSE identification. 
       The accepted events have been identified as SSE events 
	by all our three methods of pulse shape analysis
\cite{HelKK00,KKMaj99}. 
	Shown are also the lines identified  
	by the Bayesian method (see text).
\label{fig:Spectr-54_98kgy}}
\end{figure}

	The data taken in the period August 1990 - May 2000 
	(54.9813\,kg\,y, or 723.44 mol-years are shown in Fig. 
\ref{fig:Spectr-54_98kgy}. 
	Also shown in Fig. 
\ref{fig:Spectr-54_98kgy}. 
	are the data of single site events taken in the period 
	November 1995 - May 2000 with our methods of pulse shape 
	analysis (PSA) 
\cite{HelKK00,KKMaj99,Patent-KKHel}.
	We have analysed 
\cite{KK02}
	those data with various statistical methods,  
	in particular also with the  
	Bayesian method (see, e.g. 
\cite{Bayes_Method-General}).  
	This method is particularly suited for low
	counting rates, where the data follow 
	a Poisson distribution, that cannot be approximated by a Gaussian.
	
\vspace{-0.3cm}



\begin{figure}[h]
\begin{center}
\includegraphics[scale=0.2]
{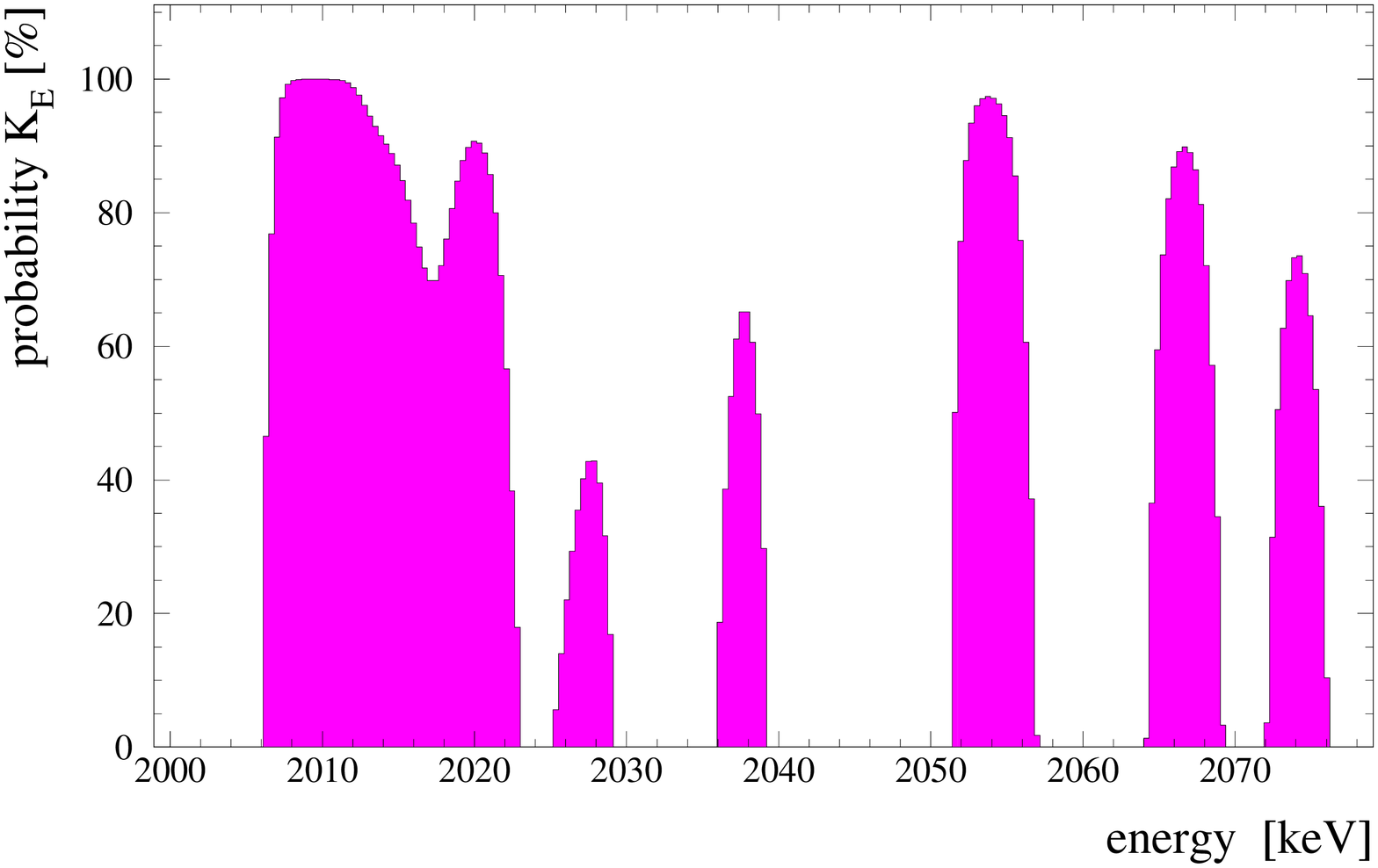} 
\hspace{.3cm}
\includegraphics[scale=0.2]
{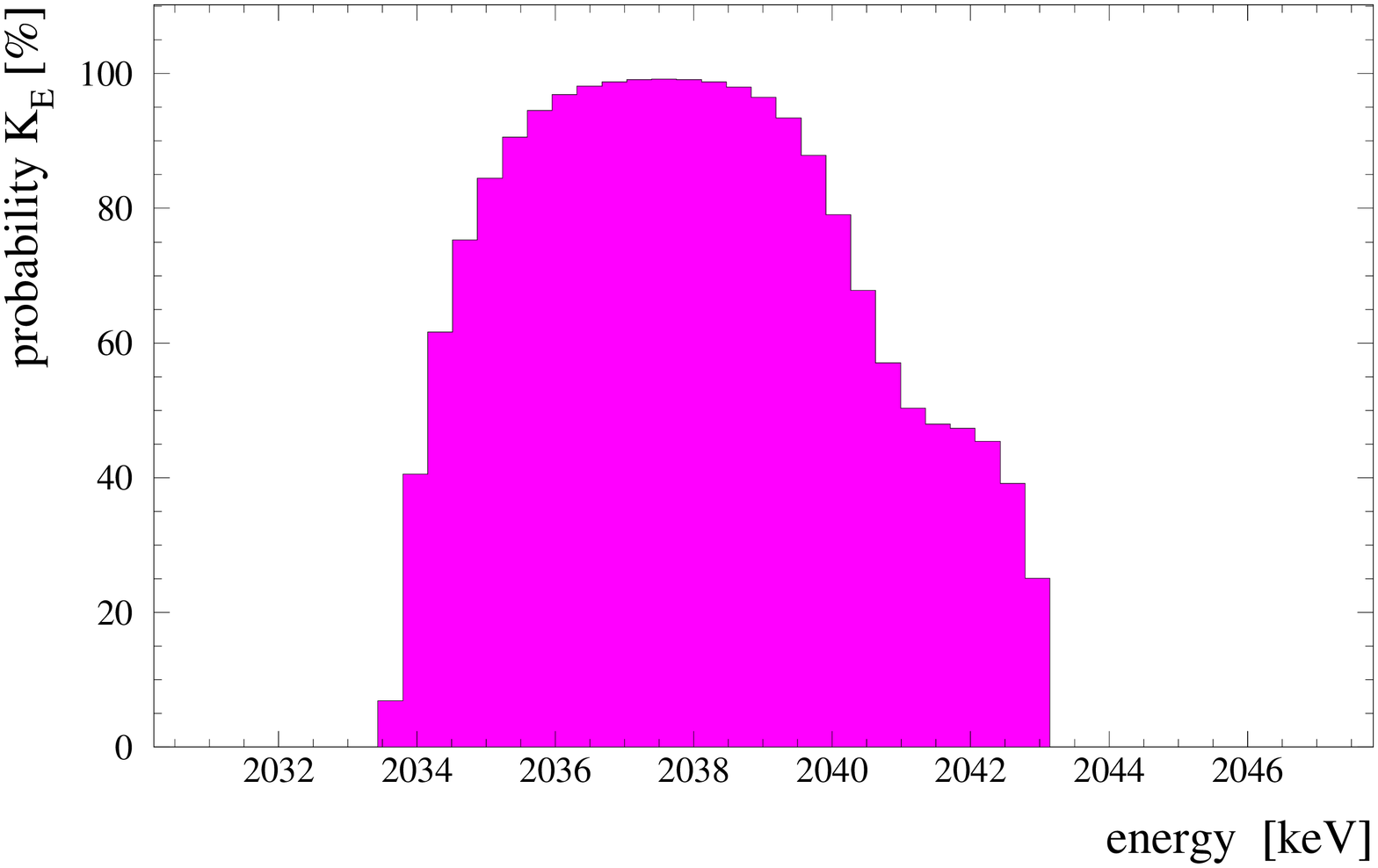} 
\end{center}

\vspace{-.7cm}
	\caption{Left: 
	Probability K that a line exists at a given energy in the 
	range of 2000-2080 keV derived via Bayesian inference 
	from the spectrum shown in Fig.
\ref{fig:Spectr-54_98kgy} (left part). 
	Right: 
	Result of a Bayesian scan for lines as in 
	the left part of this figure,  
	but in the energy range of interest for double beta decay. 
\label{fig:Bay-Chi-all-90-00-gr}}
\end{figure}



\begin{figure}[t]
\begin{center}
\includegraphics[scale=0.2]
{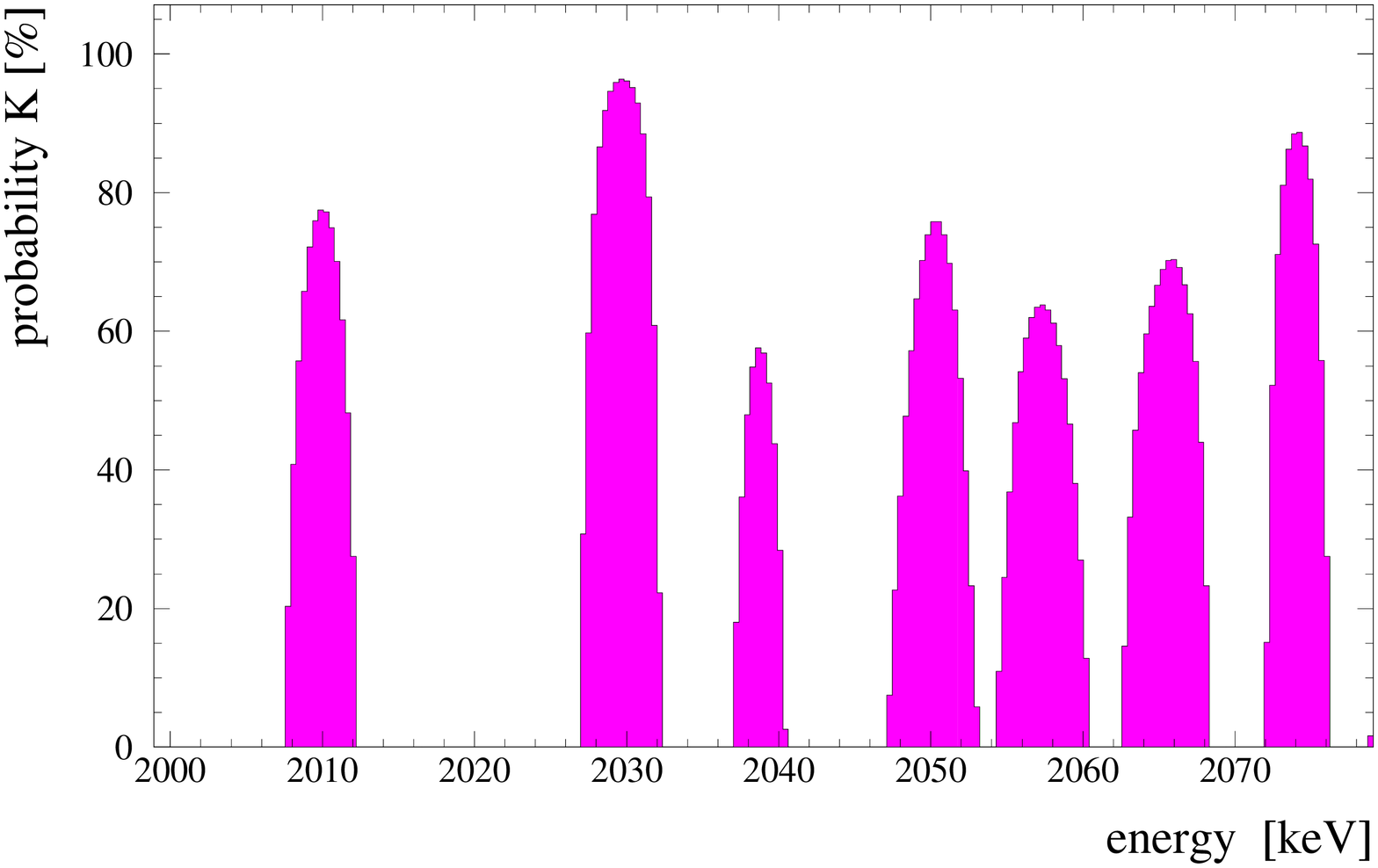} 
\hspace{.3cm}
\includegraphics[scale=0.2]
{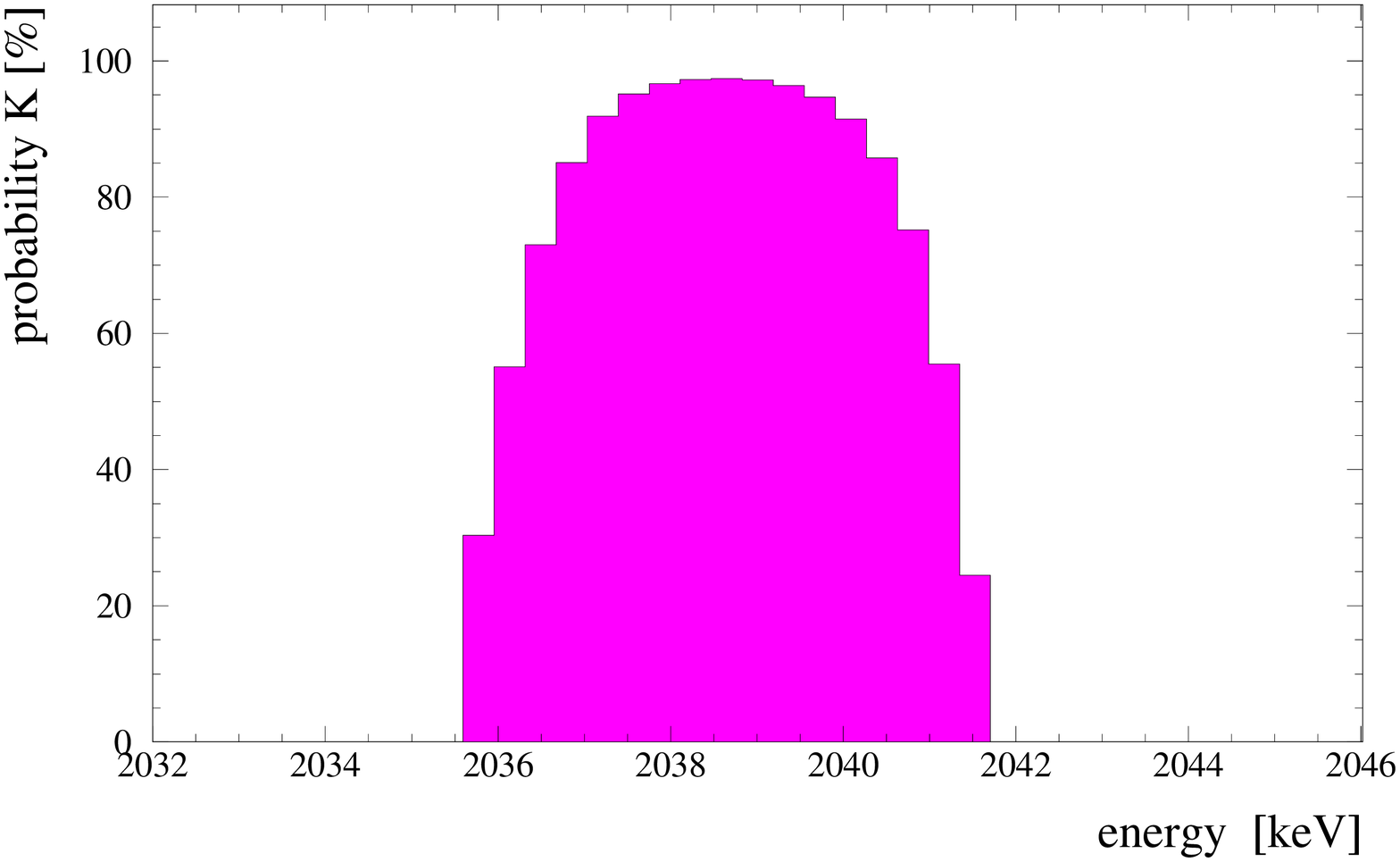} 
\end{center}

\vspace{-.7cm}
\caption{Scan for lines in the single site event spectrum 
	taken from 1995-2000 with detectors Nr. 2,3,5, 
	(Fig. 
\ref{fig:Spectr-54_98kgy}, right part), 
	with the Bayesian method. 
	Left: Energy range 2030 -2080 keV. 
	Right: Energy range of interest for double beta decay.}
\label{fig:Bay-Hell-95-00-gr-kl-Bereich}
\end{figure}


	Figs.%
~\ref{fig:Bay-Chi-all-90-00-gr},\ref{fig:Bay-Hell-95-00-gr-kl-Bereich}
	show the results of the analysis of the data with 
	the Bayesian method. Shown is the 
	probability K to find a (Gaussian) 
	line with known standard deviation of 
	$\sigma$ = 1.70 (1.59) keV corresponding to 4.00 (3.74), 
	for 3.74 keV FWHM, as function of energy. 
	This information was obtained by two-parametric Bayesian inference. 
	We have asked for the intensity I of a line at a given energy E and 
	for the background level. The background was assumed 
	to be constant over the spectral range of 2000-2080 keV. 
	For details of the procedure see 
\cite{KK02}.
	The probability K is the ordinate in Figs. 
\ref{fig:Bay-Chi-all-90-00-gr},\ref{fig:Bay-Hell-95-00-gr-kl-Bereich}.

	This procedure reproduces $\gamma$-lines 
	at the position of  known weak lines  
	from the decay of $^{214}{Bi}$ at 2010.7, 2016.7, 2021.8 
	and 2052.9 keV 
\cite{Tabl-Isot96}. 
	In addition, a line centered at 2039 keV shows up. 
	This is compatible with the Q-value 
\cite{Old-Q-val,New-Q-2001}
	of the double beta decay process. 
	We emphasize, 
	that at this energy no $\gamma$-line is expected 
	according to the compilations in 
\cite{Tabl-Isot96}. 
	The background on the left-hand side identified 
	by the Bayesian method is too high, because 
	the procedure averages the background over 
	all the spectrum (including lines), except the line 
	it is trying to single out.
	Therefore, on the right side of Fig. 
\ref{fig:Bay-Chi-all-90-00-gr} 
	we show the same as on the left side, but in the spectral range 
	of interest for double beta decay.
	Here the background is determined from a $\pm\sim$ 5 $\sigma$ 
	energy interval, around Q$_{\beta\beta}$, 
	where no lines are expected.
	The Bayesian analysis of this energy range 
	yields a confidence level (i.e. the probability K) 
	for a line to exist at 
	2039.0 keV of 96.5 $\%$ c.l. (2.1 $\sigma$).  
	We repeated the analysis for the same data, 
	but except detector 4, which had no muon shield 
	and a slightly worse energy resolution (46.502\,kg\,y). 
	The result is similar to that given in Fig. 
\ref{fig:Bay-Chi-all-90-00-gr}, 
	the probability we find for a line at 2039.0 keV 
	is 97.4$\%$ (2.2 $\sigma$) 
\cite{KK02}.

	We also applied the method recommended by the Particle Data Group 
\cite{RPD00}.
	This method 
	(which does not use the information 
	that the line is Gaussian) finds 
	a line at 2039 keV on a confidence level of 
	3.1 $\sigma$ (99.8$\%$ c.l.).

	The spectrum of single site events selects events 
	confined to a few mm region in the detector corresponding 
	to the track length of the emitted electrons - such 
	as double beta events, and rejects multiple-site events - 
	such as multiple compton scattering events 
\cite{HelKK00,Patent-KKHel,KKMaj99}.

	The expectation for a \znbb   ~~signal would be a line of single 
	site events on some background of multiple site events 
	but also single site events, the latter coming 
	e.g. from the continuum of the 2614 keV $\gamma$-line from 
	$^{208}{Tl}$ (see, e.g., the simulation in 
\cite{HelKK00}).

	Installation of PSA has been performed in 1995 for the  
	four large detectors. Detector 
	Nr.5 runs since February 1995, detectors 2,3,4 since 
	November 1995 with PSA. 
	The measuring time with PSA  
	from November 1995 until May 2000 is 36.532 kg years, 
	for detectors 2,3,5 it is 28.053\,kg\,y.

\begin{figure}[h]
\begin{center}
\includegraphics[scale=0.2]{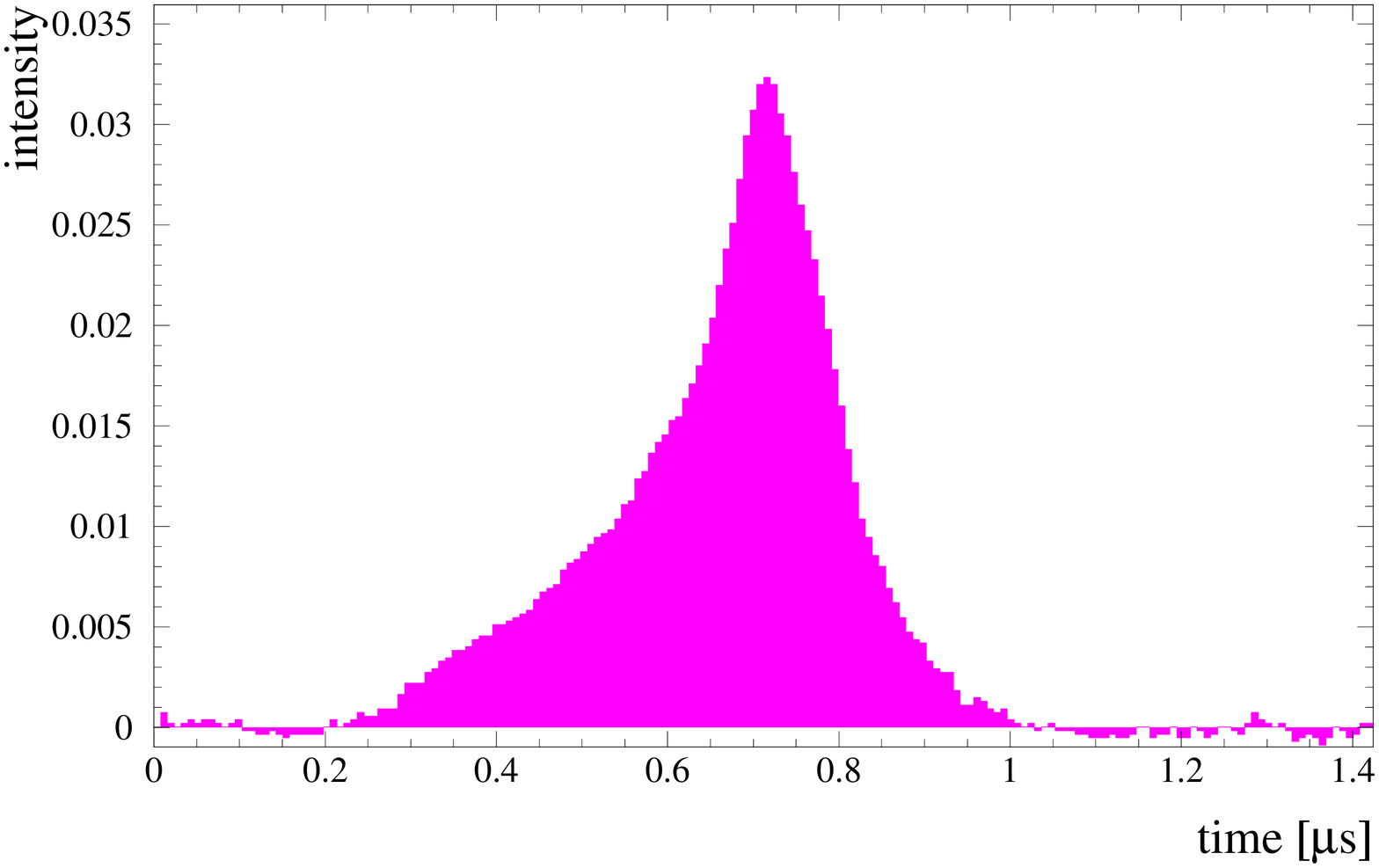} 
\hspace{.3cm}
\includegraphics[scale=0.2]{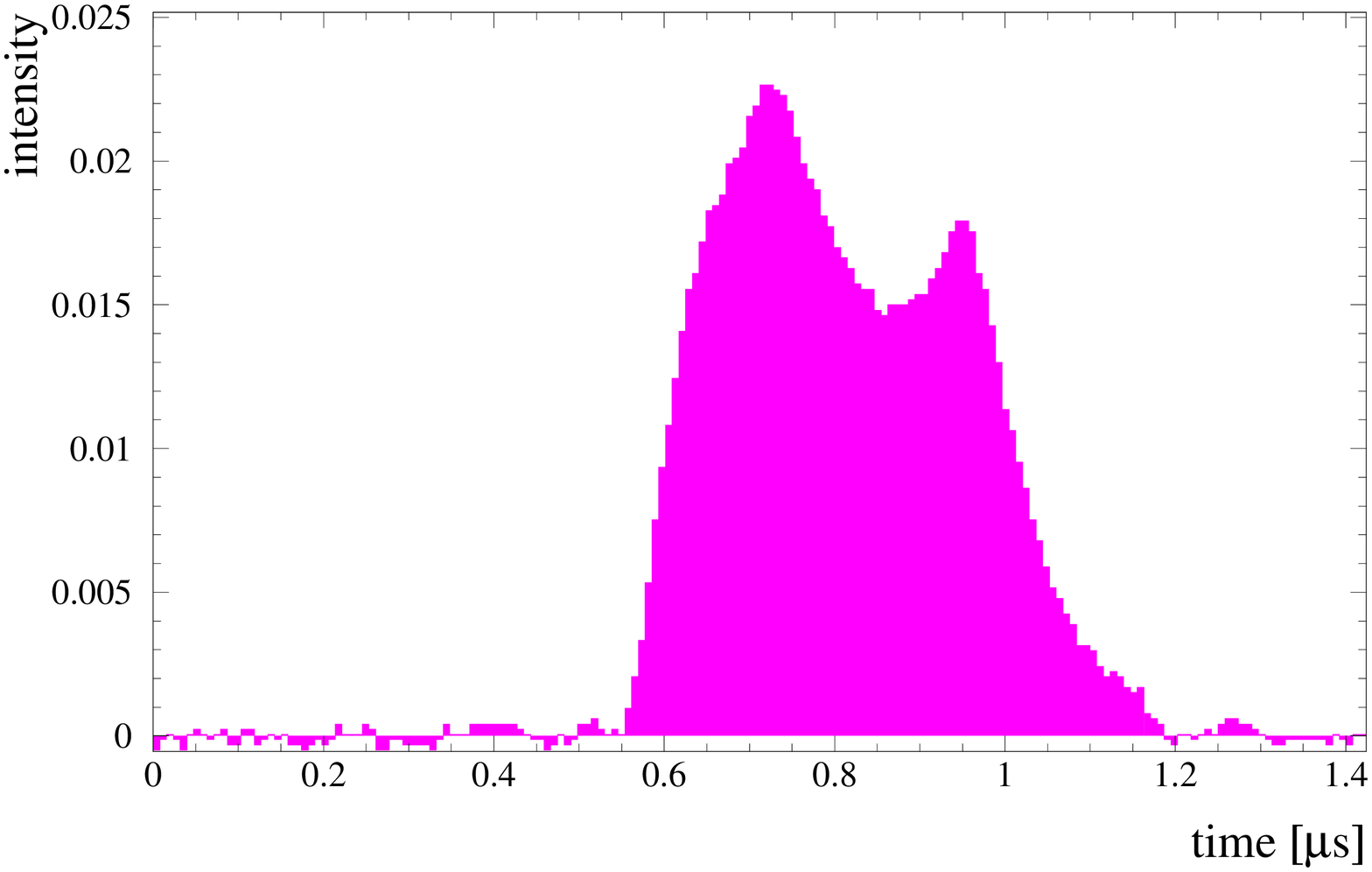} 
\end{center}

\vspace{-.7cm}
\caption{Left: Shape of one candidate for \znbb ~~decay (energy 
	2038.61\,keV) classified as SSE by all three methods 
	of pulse shape discrimination.  
	Right: Shape of one candidate (energy 
	2038.97\,keV) 
	classified as MSE by all three methods.}
\label{fig:Shape}
\end{figure}

	Fig. 
\ref{fig:Spectr-54_98kgy}. 
	shows the SSE spectrum obtained 
	under the restriction that the signal simultaneously fulfills  
	the criteria of {\it all three} methods for a single site event. 
	Fig. 
\ref{fig:Shape} 
	shows typical SSE and MSE events from our spectrum.

	In total, we find 9 SSE events in the 
	region 2034.1 - 2044.9\,keV 
	($\pm$ 3$\sigma$ around Q$_{\beta\beta}$).  
	Corrected for the efficiency to identifity an SSE signal 
	by successive application of all three PSA methods, 
	which is 0.55 $\pm$ 0.10, we obtain a \znbb~ signal of 
	(3.6 - 12.5)\,events 
	with 68.3$\%$ c.l. (best value 8.3\,events).
	Analysis with the Bayesian method 
	of the range 2000 - 2080\,keV shows again evidence for a line   
	at the energy of 2039.0\,keV 
	(Fig.  
\ref{fig:Bay-Hell-95-00-gr-kl-Bereich}).
	The analysis of the range 2032 - 2046\,keV yields 
	a signal of single site events,  
	as expected 
	for neutrinoless double beta decay, with 96.8$\%$ c.l. 
	precisely at the Q$_{\beta\beta}$ value obtained 
	in the precision experiment of 
\cite{New-Q-2001} 	
	(Fig. 
\ref{fig:Bay-Hell-95-00-gr-kl-Bereich}). 
	The PDG method gives a signal at 2039.0\,keV of 2.8 $\sigma$ 
	(99.4$\%$).


\vspace{-0.3cm}
\subsection{\it Half-Life and Effective Neutrino Mass}

	Under the assumption that the signal at Q$_{\beta\beta}$ is not 
	produced by a background line of at present unknown origin, 
	we can translate the observed number of events into half-lifes. 
	In Table 
\ref{Results} 
	we give conservatively the values obtained 
	with the Bayesian method. 
	Also given are the effective neutrino masses 
	$\langle m_\nu \rangle$ deduced using matrix elements from 
\cite{StMutKK90,Tom91}.


\begin{table}[h]

\caption[]{Half-life for the neutrinoless decay mode 
	and deduced effective neutrino mass 
	from the HEIDELBERG-MOSCOW experiment \cite{KK02}.}
\begin{center}
\newcommand{\m}{\hphantom{$-$}}
\renewcommand{\arraystretch}{1.}
\setlength\tabcolsep{5.8pt}
\begin{tabular}{c|c|c|c|c}
\hline
\hline
&&&&\\
Significan-	&	Detectors		
&	${\rm T}_{1/2}^{0\nu}	{\rm~ \,y}$	
& $\langle m \rangle $ eV	&	Conf. \\
	ce $[ kg\,y ]$	&&&& level\\
\hline
&&&&\\
	54.9813	
&	1,2,3,4,5
&	$(0.80 - 35.07) \times 10^{25}$	
& (0.08 - 0.54)		
& 	$95\%  ~c.l.$	\\
	54.9813	
&	1,2,3,4,5
&	$(1.04 - 3.46) \times 10^{25}$	
& (0.26 - 0.47)	
& $68\%  ~c.l.$	\\
 	54.9813	
&	1,2,3,4,5
&	$1.61 \times 10^{25}$	
& 0.38 
& Best Value	\\
\hline
 	46.502	
&	1,2,3,5
&	$(0.75 - 18.33) \times 10^{25}$	
& (0.11 - 0.56)
& $95\%  ~c.l.$	\\
	46.502	
&	1,2,3,5
&	$(0.98 - 3.05) \times 10^{25}$	
& (0.28 - 0.49)
& $68\% ~c.l.$	\\
	46.502	
&	1,2,3,5
&	$1.50 \times 10^{25}$	
& 0.39
& Best Value		\\
\hline
	28.053	
&	2,3,5  SSE	
&	$(0.88 - 22.38) \times 10^{25}$	
& (0.10 - 0.51)	
& $90\% ~c.l.$		\\
	28.053	
&	2,3,5  SSE	
&	$(1.07 - 3.69) \times 10^{25}$	
& (0.25 - 0.47)		
& $68\% ~c.l.$		\\
	28.053	
&	2,3,5 SSE
&	 $1.61 \times 10^{25}$	
& 0.38
& Best Value	\\
\hline
\hline
\end{tabular}
\end{center}
\label{Results}
\end{table}


	We derive from the data taken with 46.502\,kg\,y 
	the half-life 
	${\rm T}_{1/2}^{0\nu} = (0.8 - 18.3) \times 10^{25}$ 
	${\rm y}$ (95$\%$ c.l.). 
	The analysis of the other data sets, shown in Table 
\ref{Results},
	and in particular of the single site events data, 
	which play an important role in our  
	conclusion, confirm this result.

	The result obtained is consistent with the limits 
	given earlier by the HEI\-DELBERG-MOSCOW experiment 
\cite{HDM01}, 
	considering that the background had been determined 
	more conservatively there.
	It is also consistent with all other double beta experiments -  
	which still reach less sensitivity. 
	A second Ge-expe\-riment, 
	which has stopped operation in 1999 after 
	reaching a significance of 9\,kg\,y,
	yields (if one, believes their method of 'visual inspection' 
	in their data analysis), in a conservative analysis, a limit of 
	${\rm T}_{1/2}^{0\nu} > 0.55 \times 10^{25}
	{\rm~ y}$  (90\% c.l.). 
	The $^{128}{Te}$ geochemical experiment 
	yields $\langle m_\nu \rangle < 1.1$ eV (68 $\%$ c.l.), 
	the $^{130}{Te}$ cryogenic experiment yields 
	$\langle m_\nu \rangle < 1.8$\,eV 
	and the CdWO$_4$ experiment $\langle m_\nu \rangle < 2.6$\,eV,  
	all derived with the matrix elements of 
\cite{Sta90} 
	to make the results comparable to the present value 
	(for references see 
\cite{KK60Y}).

	Concluding we obtain, with about 95$\%$ probability, 
	first evidence for the neutrinoless 
	double beta decay mode. 
	As a consequence, at this confidence level, 
	lepton number is not conserved. 
	Further the neutrino is a Majorana particle. 
	The effective mass 
	$\langle m \rangle $ is deduced 
	to be $\langle m \rangle $ 
	= (0.11 - 0.56)\,eV (95$\%$ c.l.), 
	with best value of 0.39\,eV. 
	Allowing conservatively for an uncertainty of the nuclear 
	matrix elements of $\pm$ 50$\%$
	(for detailed discussions of the status 
	of nuclear matrix elements we refer to 
\cite{KK60Y,Tom91} 
	and references therein)
	this range may widen to 
	$\langle m \rangle $ 
	= (0.05 - 0.84)\,eV (95$\%$ c.l.). 

	In this conclusion, it is assumed that contributions 
	to \znbb~ decay from processes other than  
	the exchange of 
	a Majorana neutrino (see, e.g. 
\cite{KK60Y}
	and references therein) are negligible.

\begin{figure}[t]
\centering{
\includegraphics*[scale=0.35]
{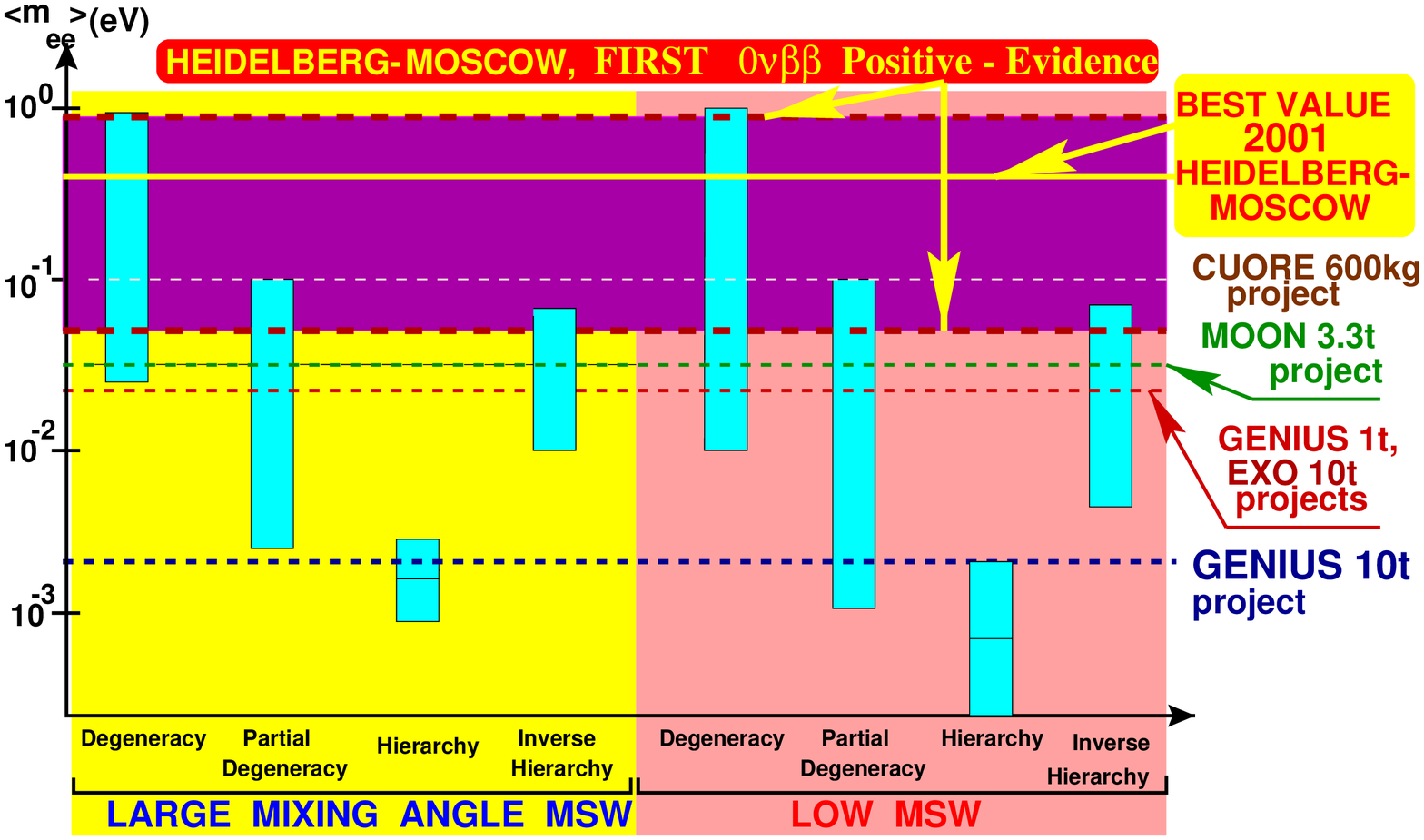}}
\caption[]{
	The impact of the evidence obtained for neutrinoless 
	double beta decay in this paper (best value 
	of the effective neutrino mass 
	$\langle m \rangle$ = 0.39\,eV, 95$\%$ 
	confidence range (0.05 - 0.84)\,eV - 
	allowing already for an uncertainty of the nuclear 
	matrix element of a factor of $\pm$ 50$\%$) 
	on possible neutrino mass schemes. 
	The bars denote allowed ranges of $\langle m \rangle$ 
	in different neutrino mass scenarios, 
	still allowed by neutrino oscillation experiments (see 
\cite{KKPS,KKPS-01}). 
	Hierarchical models are excluded by the 
	new \znbb ~~decay result. Also shown are 
	the expected sensitivities 
	for the future potential double beta experiments 
	CUORE, MOON, EXO  
	and the 1 ton and 10 ton project of GENIUS 
\cite{KK60Y,KK-SprTracts00,KK-00-NOON-NOW-NANP-Bey97-GEN-prop}.
\label{fig:Jahr00-Sum-difSchemNeutr}}
%
\centering{
\includegraphics*[scale=0.30]
{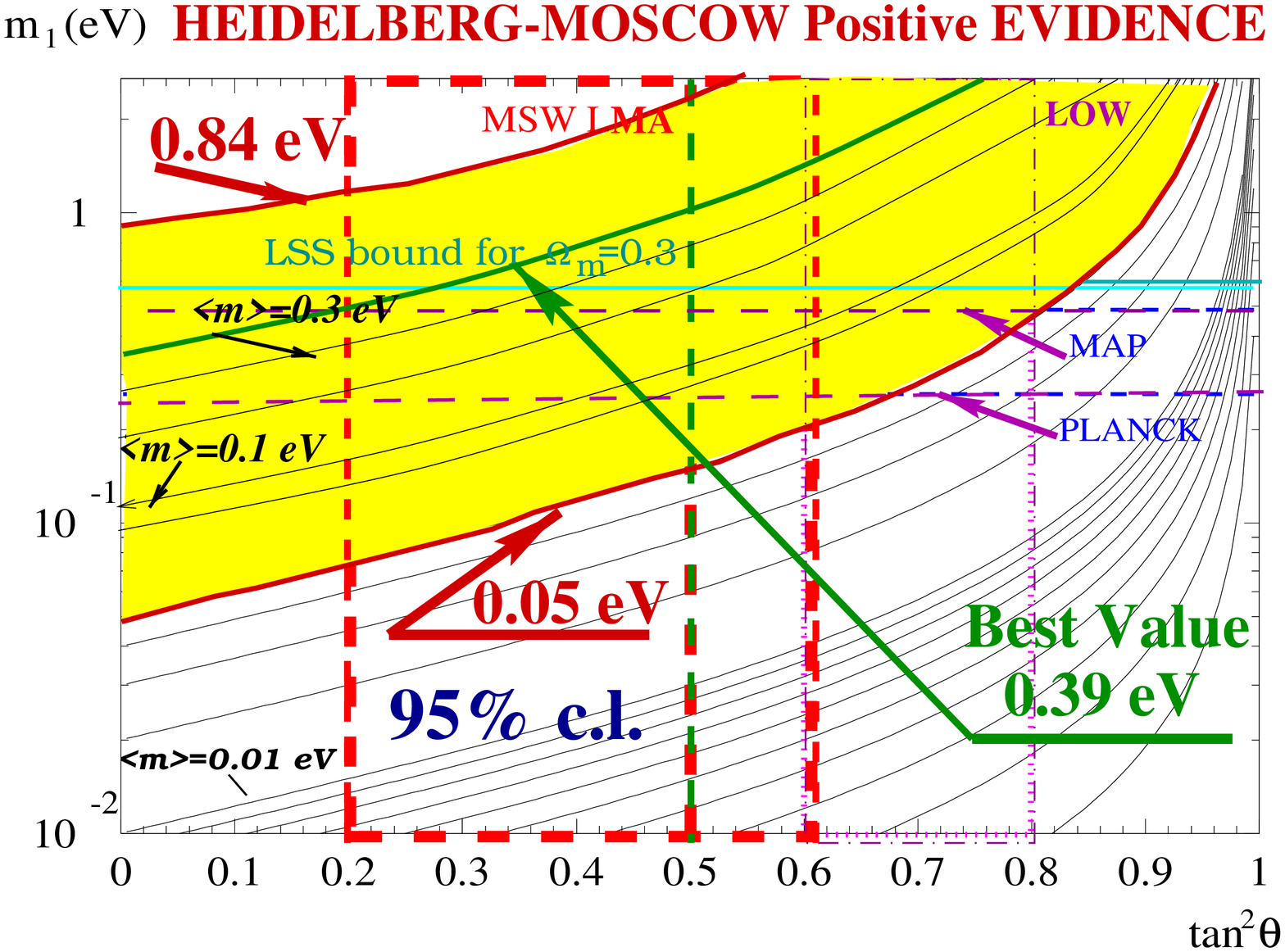}}

\vspace{-0.3cm}
\caption[]{
       Double beta decay observable 
$\langle m\rangle$
	 and oscillation parameters: 
	 The case for degenerate neutrinos. 
	 Plotted on the axes are the overall scale of neutrino masses 
	 $m_0$ and mixing $\tan^2\, \theta^{}_{12}$. 
	 Also shown is a cosmological bound deduced from a fit of 
	 CMB and large scale structure 
\cite{Lop} 
	and the expected sensitivity of the satellite experiments 
	 MAP and PLANCK. 
	 The present limit from tritium $\beta$ decay of 2.2\,eV 
\cite{Weinh-Neu00} 
	would lie near the top of the figure. 
     The range of 
$\langle m\rangle$
	 fixed by the HEIDELBERG-MOSCOW experiment 
\cite{KK02}
	is, in the case of small solar neutrino mixing, already in the 
	 range to be explored by MAP and PLANCK 
\cite{Lop}.
\label{fig:Dark3}}
\end{figure}

	With the limit deduced for the effective neutrino mass,  
	the HEIDELBERG-MOSCOW experiment excludes several 
	of the neutrino mass 
	scenarios 
	allowed from present neutrino oscillation experiments
	(see Fig.
\ref{fig:Jahr00-Sum-difSchemNeutr}) 
	- allowing mainly only for degenerate 
	and partially degenerate mass 
	scenarios and an inverse hierarchy 3$\nu$ - scenario
	(the latter being, however, strongly disfavored 
	by a recent analysis
	of SN1987A.   
	In particular hierarchical mass schemes 
	are excluded at the present level of accuracy.

	Assuming the degenerate scenarios to be realized in nature 
	we fix - according to the formulae derived in 
\cite{KKPS} - 
	the common mass eigenvalue of the degenerate neutrinos 
	to m = (0.05 - 3.4)\,eV. 
	Part of the upper range is already excluded by 
	tritium experiments, which give a limit of m $<$ 2.2\,eV (95$\%$ c.l.) 
\cite{Weinh-Neu00}.
	The full range can only  partly 
	(down to $\sim$ 0.5\,eV) be checked by future  
	tritium decay experiments,  
	but could be checked by some future $\beta\beta$

\clearpage	


\clearpage
\noindent
	experiments (see, e.g. next section).
	The deduced best value for the mass 
	is consistent with expectations from experimental 
	$\mu ~\to~ e\gamma$
	branching limits in models assuming the generating 
	mechanism for the neutrino mass to be also responsible 
	for the recent indication for as anomalous magnetic moment 
	of the muon
\cite{MaRaid01}.
	It lies in a range of interest also for Z-burst models recently 
	discussed as explanation for super-high energy cosmic ray events 
	beyond the GKZ-cutoff 
\cite{PW01-Wail99}.
	The range of $\langle m \rangle $ fixed in this 
	work is, already now, in the range to be explored 
	by the satellite experiments MAP and PLANCK 
\cite{Lop}
	(see Fig. 
\ref{fig:Dark3}).	

	The neutrino mass deduced allows neutrinos to still play 
	an important role as hot dark matter in the Universe.


\section{\boldmath Future of $\beta\beta$ Experiments}

	To improve the present sensitivity for the effective neutrino mass 
	considerably, 
	and to fix this quantity more accurately,  
		   requires new experimental approaches, 
		   as discussed extensively in  
\cite{KK60Y,KK-NANPino00,KK-NOW00,KK-Bey97,GEN-prop,KK-Neutr98}. 
	Some of them are indicated in Figs. 
\ref{fig:Now4-gist-mass},\ref{fig:Jahr00-Sum-difSchemNeutr}.

      	It has been discussed earlier (see e.g.  
\cite{KK60Y,KK-NANPino00,KK-NOW00,KK-Bey97,KK-Neutr98,KK-NOW00}),
	that of present generation experiments probably no one has a 
	potential to probe   
$\langle m\rangle$
	below the present HEIDELBERG-MOSCOW level (see 
Fig.~\ref{fig:Now4-gist-mass}).

	 The Milano cryogenic experiment using TeO$_2$ bolometers 
	 improved their values for the 
$\langle m \rangle$ 
	 from $\beta\beta$ decay of $^{130}$Te, from 5.3 eV in 1994    
	to 1.8 eV in 2000   
\cite{Ales00}.
	NEMO-III, originally aiming at a sensitivity of 0.1 eV, 
	reduced their goals recently to $0.3\div0.7$~eV (see  
\cite{NEMO-Neutr00}
	) (which is more consistent with estimates given by   
\cite{Tret95}
	), to be reached in 6 years from starting of running, 
       foreseen for the year 2002.


\subsection{\it GENIUS and other Proposed Future Double Beta Experiments}

	With the era of the HEIDELBERG-MOSCOW experiment 
	the time of the small smart 
	experiments is over.

	To reach 
	significantly larger sensitivity, $\beta\beta$ 
	     experiments have to become large. 
	     On the other hand source strengths of up to 10 tons of 
	     enriched material touch the world production limits. 
	This means that the background has to be reduced by the order a 
	     factor of 1000 and more compared to that 
	     of the HEIDELBERG-MOSCOW experiment.

	Table 
\ref{List-exp} 
	lists some key numbers for GENIUS,  
\cite{KK-Bey97,GEN-prop,KK-NOW00} 
	     which was the 
	     first proposal for a third generation double beta experiment, 
	and of some other 
	     proposals made after the GENIUS proposal. 
	     The potential of some of them is shown also in 
Fig.~\ref{fig:Jahr00-Sum-difSchemNeutr}, and 
	it is seen that not all of them will lead to large 
	improvements in sensitivity.
	Among the latter is also the recently presented MAJORANA project 
\cite{MAJOR-WIPP00}, 
	which does not really apply a striking new strategy 
	for background reduction, 
	particularly also after it was found that the projected  
	segmentation of detectors may not work. 


\begin{table*}[t]
\caption{Some key numbers of future double beta decay experiments (and of 
	the {\sf HEIDELBERG-MOSCOW} experiment). Explanations: 
	${\nabla}$ - assuming the background of the present pilot project. 
	$\ast\ast$ - with matrix element from  
\protect\cite{StMutKK90}, 
\protect\cite{Tom91}, 
\protect\cite{Hax84}, 
\protect\cite{WuStKKChTs91}, 
\protect\cite{WuStKuKK92} (see Table II in 
\protect\cite{HM99} 
	and {\it including} an assumed uncertainty 
	of $\pm$50$\%$ of the nuclear matrix element). 
	${\triangle}$ - this case shown 
	to demonstrate {\bf the ultimate limit} of such experiments. 
	For details see 
\protect\cite{KK60Y}.}
\label{table:1}
\newcommand{\m}{\hphantom{$-$}}
\newcommand{\cc}[1]{\multicolumn{1}{c}{#1}}
\renewcommand{\arraystretch}{.2}
\setlength\tabcolsep{.9pt}
{\footnotesize
{
\begin{center}  
\begin{tabular}[!h]{|c|c|c|c|c|c|c|c|}
\hline
\hline
 &  &  &  & Assumed &  &  & \\
 &  &  &  & backgr. & $Running$ & Results & \\
$\beta\beta$-- & & & Mass & $\dag$ events/ & 	 & limit for & 
${<}m_{\nu}{>}$ \\
$Isoto-$ & $Name$ & $Status$ & $(ton-$ & kg y keV, & Time  
& $0\nu\beta\beta$ & \\
pe & & & $nes)$ & $\ddag$ events/kg & 	 & half-life & ( eV )\\ 
& & & & y FWHM,  & (tonn. & (years) & \\
& & & & $\ast$ events & years) &  & \\
& & &  & /yFWHM &  &  & \\
\hline
\hline
 &  &  &  &  &  &  & \\
~${\bf ^{76}{Ge}}$ & {\bf HEIDEL-} & {\bf run-}  & 0.011 & $\dag$ 0.06 
& {\bf 54.98} & ${\bf 0.8-18.3}$ 
& {\bf 0.05-0.84}\\
 & {\bf BERG}  & {\bf ning}  &  (enri-  &  &  {\bf kgy} 
&  ${\bf x~{10}^{25} y}$ 	& 	eV~~$)^{\ast\ast}$\\
& {\bf MOSCOW} & {\bf since} & ched) & $\ddag$ 0.24  &  
& {\bf 95$\%$ c.l.}	& {\bf 95$\%$ c.l.} \\
& {\bf \cite{KK02,HDM01,KK-SprTracts00}} & {\bf 1990} &  & $\ast$ 2 & & 
{\bf NOW !!}  &  {\bf NOW !!}\\
\hline
\hline
\hline
 &  &  &  &  &  &  & \\
${\bf ^{100}{Mo}}$ & {\sf NEMO III} & {\it under} & $\sim$0.01 & $\dag$ 
{\bf 0.0005} &  &  &\\
 & {\tt \cite{NEMO-Neutr00}}& {\it constr.} & (enri- & $\ddag$ 0.2  & 50 & 
${10}^{24}$ & 0.3-0.7\\
 &  & {\it end 2001?} & -ched) &  $\ast$ 2 &kg y  &  &\\
\hline
\hline
&  &  &  &  &  &  & \\
${\bf ^{130}{Te}}$ & ${\sf CUORE}^{\nabla}$ & {\it idea} & 0.75 & $\dag$ 0.5 
& 5 & $9\cdot{10}^{24}$ & 0.2-0.5\\
 & {\tt \cite{CUORE-LeptBar98}}& {\it since 1998} &(nat.)  
& $\ddag$ 4.5/$\ast$ 1000 &  & & \\ 
\hline
&  &  &  &  &  &  &  \\
${\bf ^{130}{Te}}$ & {\sf CUORE}  &  {\it idea} & 0.75   & $\dag$ 0.005 & 5 
& $9\cdot{10}^{25}$ & 0.07-0.2\\
&  {\tt \cite{CUORE-LeptBar98,Fior-Neutr00}} & {\it since 1998}   & (nat.)  
&  $\ddag$ 0.045/ $\ast$ 45 &  & &\\
\hline
&  &  &  &  &  &  & \\
${\bf ^{100}{Mo}}$ & {\sf MOON} & {\it idea} & 10 (enr.) & ? & 30 & ? &\\
 & {\tt \cite{Ej00,LowNu2}} & {\it since 1999} &  100(nat.) & & 300 & &0.03 \\
\hline
&  &  &  &  &  &  & \\
${\bf ^{116}{Cd}}$ & {\sf CAMEOII} & {\it idea}  & 0.65 & * 3. & 5-8  
& ${10}^{26}$ & 0.06 \\
& {\sf CAMEOIII}{\sf \cite{Bell00}} & {\it since 2000 } & 1(enr.) & ? & 5-8 
&  ${10}^{27}$ & 0.02 \\
\hline
&  &  &  &  &  &  & \\
${\bf ^{136}{Xe}}$ & {\sf EXO} & Proposal& 1 & $\ast$ 0.4 & 5 & 
$8.3\cdot{10}^{26}$ & 0.05-0.14\\
&  & since &  &  &  &  & \\
  & {\tt \cite{EXO00,EXO-LowNu2}} & 1999 & 10 & $\ast$ 0.6 & 10 & 
$1.3\cdot{10}^{28}$ & 0.01-0.04\\
\hline 
\hline
\hline
\hline
&  &  &  &  &  &  &  \\
~${\bf ^{76}{Ge}}$ & {\bf GENIUS} & {\it under} & 11 kg & 
$\dag$ ${\bf 6\cdot{10}^{-3}}$& 3 
& {\bf ${\bf 1.6\cdot{10}^{26}}$} & {\bf 0.15} \\
& {\bf - TF} & {\it constr.} & (enr.) &  &  &  &   \\
&  {\bf \cite{KK-GeTF-MPI,GenTF-0012022}}&  {\it end 2001?} &  & &  &  &  \\
\hline
&  &  &  &  &  &  &  \\
~${\bf ^{76}{Ge}}$ & {\bf GENIUS} & Pro- & 1  & $\dag$ 
${\bf 0.04\cdot{10}^{-3}}$ & 1 & ${\bf 5.8\cdot{10}^{27}}$ & 
{\bf 0.02-0.05} \\
 & {\tt \cite{KK-Bey97,GEN-prop}}  & posal &(enr.)  
& $\ddag$ ${\bf 0.15\cdot{10}^{-3}}$ & & & \\
&  & since &  & $\ast$ {\bf 0.15} &  &  &  \\
&  & 1997 & 1 & ${\bf \ast~ 1.5}$ & 10 & ${\bf 2\cdot{10}^{28}}$  & 
{\bf 0.01-0.028} \\
\hline
&  &  &  &  &  &  &  \\
~${\bf ^{76}{Ge}}$ & {\bf GENIUS} & Pro- & 10 
& $\ddag$ ${\bf 0.15\cdot{10}^{-3}}$ & 10 &
${\bf 6\cdot{10}^{28}}$ & {\bf 0.006 -}\\
&  {\tt \cite{KK-Bey97,GEN-prop}} &  posal &  &  &  &  &  {\bf 0.016}\\
 &   &  since &(enr.) & ${\bf 0^{\triangle}}$ & 10 & 
${\bf 5.7\cdot{10}^{29}}$ & {\bf 0.002 -}\\
&  &  1997 &  &  &  &  &  {\bf 0.0056}\\ 
\hline 
\hline
\end{tabular}\\[2pt]
\end{center}}}
\label{List-exp}
\end{table*}


	For more recent information on XMASS, EXO, MOON experiments see 
	the contributions of 
	Y. Suzuki, G. Gratta and H. Ejiri in Ref. 
\cite{LowNu2}.
	     The CAMEO project 
\cite{Bell00}
	in its now propagated variant GEM  
	is nothing then 
	a variant of GENIUS (see below) put into the BOREXINO tank, 
	at some later time. 	
	CUORE  
\cite{CUORE-LeptBar98} 
	has, with the complexity of cryogenic techniques, 
	   still to overcome serious 
	   problems of background to enter 
	   into interesting regions of
$\langle m \rangle$.
	 EXO  
\cite{EXO00} 
	needs still very extensive research and development to probe 
	 the applicability of the proposed detection method.
	In particular if it would be confirmed that tracks will 
	be too short to be identified, it would act essentially 
	only as a highly complicated calorimeter.
	 In the GENIUS project a reduction by a factor of more than 1000 
	 down to a background level of 0.1 events/tonne y keV 
	 in the range of $0\nu\beta\beta$ decay is 
	planned to be reached by removing all 
	 material close to the detectors, and by using naked Germanium 
	 detectors in a large tank of liquid nitrogen. 
	 It has been shown that the detectors show excellent 
	 performance under such conditions  
\cite{GEN-prop,KK-J-PhysG98}.
	For technical questions and extensive Monte Carlo simulations of 
	the GENIUS project for its application in double beta decay 
	we refer to  
\cite{GEN-prop,KK-J-PhysG98}.

\begin{table*}[h]
\begin{center}
\caption{\label{New-Proj}Some of the new projects under discussion for future double beta 
	decay experiments (see ref.%
\protect\cite{KK60Y}).}
\label{tableA}
\newcommand{\m}{\hphantom{$-$}}
\newcommand{\cc}[1]{\multicolumn{1}{c}{#1}}
\renewcommand{\tabcolsep}{.19 pc} 
\renewcommand{\arraystretch}{.05} 
\begin{tabular}[!h]{|c|c|c|c|c|}
\hline
\hline
\multicolumn{5}{|c|}{}\\
\multicolumn{5}{|c|}{NEW~~~  PROJECTS}\\
\multicolumn{5}{|c|}{}\\
\hline
 &  &  &  & \\
 & BACKGROUND & MASS & POTENTIAL & POTENTIAL \\
 &  &  &  & \\
 & REDUCTION & INCREASE & FOR DARK & FOR SOLAR\\
 &  &  & MATTER & ${\nu}^{'}$ s\\
\hline
&  &  &  & \\
 {\bf GENIUS} 	& {\bf +} 	& {\bf +} 	
& {\bf +} & {\bf + ${}^{*)}$	}\\
\hline
&  &  &  & \\
 {\sf CUORE}  	& {\bf (+)} 	& {\bf + } 	
& {\bf $-$} & {\bf $-$	}	\\
\hline
&  &  &  & \\
 {\sf MOON}  		& {\bf (+) }	& {\bf + } 	
& {\bf $-$} & {\bf +}		\\
\hline
&  &  &  & \\
 {\bf EXO} 		&  {\bf +} 	& {\bf +}  	
& {\bf $-$} & {\bf $-$	}	\\
\hline
&  &  &  & \\
 {\sf MAJORANA}	& {\bf $-$ }	& {\bf + } 	
& {\bf $-$} & {\bf $-$	}	\\
\hline
\multicolumn{5}{|c|}{}\\
\multicolumn{5}{|c|}{\sf *) real time measurement of pp neutrinos 
	with threshold of 20 keV (!!)}\\
\multicolumn{5}{|c|}{}\\
\hline
\hline
\end{tabular}\\[1pt]
\end{center}
\end{table*}


\vspace{-0.5cm}
\subsection{\it GENIUS and Other Beyond Standard Model Physics}

		   GENIUS will allow besides the 
	large increase in sensitivity for double beta decay 
	described above, the access to a broad range 
		   of other beyond SM physics topics in the multi-TeV range. 
	Already now $\beta\beta$ decay probes the TeV scale on which new 
	physics should manifest itself (see, e.g. 
\cite{KK-Bey97,KK-LeptBar98,KK-SprTracts00}). 
	  Basing to a large extent on the theoretical work of the 
	  Heidelberg group in the last six years, the 
	  HEIDELBERG-MOSCOW experiment yields results for SUSY models 
	  (R-parity breaking, neutrino mass), leptoquarks 
	  (leptoquarks-Higgs coupling), compositeness, right-handed $W$ mass, 
	  nonconservation of Lorentz invariance and 
	  equivalence principle, mass of a heavy left or 
	  righthanded neutrino, competitive to corresponding results 
	  from high-energy accelerators like TEVATRON and HERA. 
	  The potential of GENIUS extends into the multi-TeV region for 
	  these fields and its sensitivity would correspond to that of 
	  LHC or NLC and beyond (for details see  
\cite{KK60Y,KK-LeptBar98,KK-SprTracts00}).


\subsection{\it GENIUS-Test Facility}
		     Construction of a test facility for 
		     GENIUS -- GENIUS-TF -- 
		     consisting of $\sim$~40 kg of HP Ge 
		     detectors suspended in a liquid nitrogen 
		     box has been started. 
		     Up to summer of  2001, six detectors each 
		     of $\sim$~2.5\,kg and with a threshold of as low 
		     as $\sim$~500\,eV have been produced.

         Besides test of various parameters of the GENIUS project, 
	 the test facility would allow, with the projected background 
	 of 2-4 events/(kg\,y\,keV) in the low-energy range, 
	to probe the DAMA 
	 evidence for dark matter by the seasonal modulation signature, 
	(see 
\cite{KK-GeTF-MPI,GenTF-0012022}).


\vspace{-0.3cm}
\section{Conclusion}

	The status of present double beta decay search 
	has been discussed, and 	
	recent evidence for a non-vanishing 
	Majorana neutrino mass obtained by the HEIDELBERG-MOSCOW 
	experiment has been presented. Future projects to improve 
	the present accuracy of the effective neutrino mass have 
	been discussed. The most sensitive of them and perhaps  
	at the same time most realistic one, is the GENIUS project.
	 GENIUS is the only of the new projects (see also Table 
\ref{New-Proj}) 
	which simultaneously has a huge potential for 
      cold dark matter search, and for real-time detection of 
      low-energy neutrinos (see 
\cite{KK-Bey97,KK-NOW00,BedKK-01,KK-SprTracts00,LowNu2,KK-NANPino00}).


\small{
        
}

\end{document}